\documentclass[reprint,aps,prb,showpacs,superscriptaddress,letterpaper]{revtex4-1}
\usepackage [utf8]{inputenc}
\usepackage{amsmath, nicefrac, ulem}

\usepackage{graphicx}

\newcommand{\ie}{\textit{i.\,e.}}

\newcommand{\ket}[1]{\left|#1\right\rangle}

\newcommand{\up}{\uparrow}
\newcommand{\dn}{\downarrow}
\newcommand{\ra}{\rightarrow}

\newcommand{\tip}{{\rm tip}}
\newcommand{\sub}{{\rm sub}}
\newcommand{\eff}{{\rm eff}}
\newcommand{\bg}{{\rm bg}}

\begin{document}

\title{Spin dependent transmission of nickelocene-copper contacts probed with shot noise}

\author{Michael~Mohr}
\affiliation{Institut f\"ur Experimentelle und Angewandte Physik, Christian-Albrechts-Universit\"at zu Kiel, D-24098 Kiel, Germany}
\author{Manuel~Gruber}
\affiliation{Institut f\"ur Experimentelle und Angewandte Physik, Christian-Albrechts-Universit\"at zu Kiel, D-24098 Kiel, Germany}
\author{Alexander~Weismann}
\affiliation{Institut f\"ur Experimentelle und Angewandte Physik, Christian-Albrechts-Universit\"at zu Kiel, D-24098 Kiel, Germany}
\author{David~Jacob}
\affiliation{Departamento de F\'isica de Materiales, Universidad del Pa\'is Vasco UPV/EHU, E-20018 San Sebasti\'an, Spain}
\affiliation{IKERBASQUE, Basque Foundation for Science, E-48013 Bilbao, Spain}
\author{Paula~Abufager}
\affiliation{Instituto de F\'isica de Rosario, Consejo Nacional de Investigaciones Cient\'ficas y T\'ecnicas (CONICET)}
\affiliation{Universidad Nacional de Rosario, 2000 Rosario, Argentina}
\author{Nicol\'as~Lorente}
\affiliation{Centro de F\'isica de Materiales CFM/MPC (CSIC-UPV/EHU), E-20018 Donostia-Sebasti\'an, Spain}
\affiliation{Donostia International Physics Center (DIPC), E-20018 Donostia-Sebasti\'an, Spain}
\author{Richard~Berndt}
\affiliation{Institut f\"ur Experimentelle und Angewandte Physik, Christian-Albrechts-Universit\"at zu Kiel, D-24098 Kiel, Germany}

\begin{abstract}
The current $I$ through nickelocene molecules and its noise are measured with a low temperature scanning tunneling microscope on a Cu(100) substrate.
Density functional theory calculations and many-body modeling are used to analyze the data.
During contact formation, two types of current evolution are observed, an abrupt jump to contact and a smooth transition.
These data along with conductance spectra ($dI/dV$) recorded deep in the contact range are interpreted in terms of a transition from a spin-1 to a spin-\nicefrac{1}{2} state that is Kondo screened.
Many-body calculations show that the smooth transition is also consistent with a renormalization of spin excitations of a spin-1 molecule by Kondo exchange coupling.
The shot noise is significantly reduced compared to the Schottky value of $2eI$.
The noise can be described in the Landauer picture in terms of the spin polarization of the transmission of $\approx35$\,\% through two degenerate $d_\pi$-orbitals of the Nickelocene molecule.
\end{abstract}

\pacs{
72.25.Mk, % Spin transport through interfaces
72.10.Fk, %	Scattering by point defects, dislocations, surfaces, and other imperfections (including Kondo effect)
72.70.+m,	% Noise processes and phenomena
74.55.+v, % Tunneling phenomena: single particle tunneling and STM
72.70.+m, % Noise processes and phenomena
73.63.Rt, %	Nanoscale contacts
73.50.Td  % Noise processes and phenomena
}

\maketitle

\section{Introduction}

An electrical current $I$ exhibits shot noise because the electron charge $e$ is quantized \cite{Schottky}.
While the current itself may be understood as a single-electron quantity, the shot noise reflects electron correlations and, therefore, is affected by the
Pauli principle.
As a consequence, electrons with identical spins avoid bunching, which reduces the noise.

While noise in mesoscopic structures has been investigated in some detail \cite{blanter}, noise spectroscopy of atomic \cite{VanDenBrom_PRL, RBG_Kumar_PRL, Schoenenberger, Scheer_Ir, kemiktarak_radio-frequency_2007, chen_enhanced_2014, Urbina_Al, Vardimon, vardimon2015indication, vardimon2016orbital, MKu, nonoise} and molecular \cite{RBG_Djukic_single_molecule, chain, Ruitenbeek_Benzene, karimi2016shot, talwasser, c60, totanoise} junctions is a more recent development.
No data have been reported from molecules that are likely to exhibit spin-dependent transport.
Fairly simple magnetic complexes such as metallocenes, \ie\ a transition metal or lanthanide ion sandwiched between cyclopentadienyl rings, appear to be suitable for this purpose \cite{rin11, lan13, zha14, kan16, har17, kna17, guo17, che17, Knaak2017a}.
In addition, it is advantageous to use junctions with conductances $G \sim  {\mathrm G_0}$ ($ {\mathrm G_0} = 2 e^2/h$) because the spin-related noise reduction tends to scale with the conductance~as detailed below (Eq.~\ref{twoch}).

\begin{figure}[hbt]
\includegraphics[width=0.3\columnwidth]{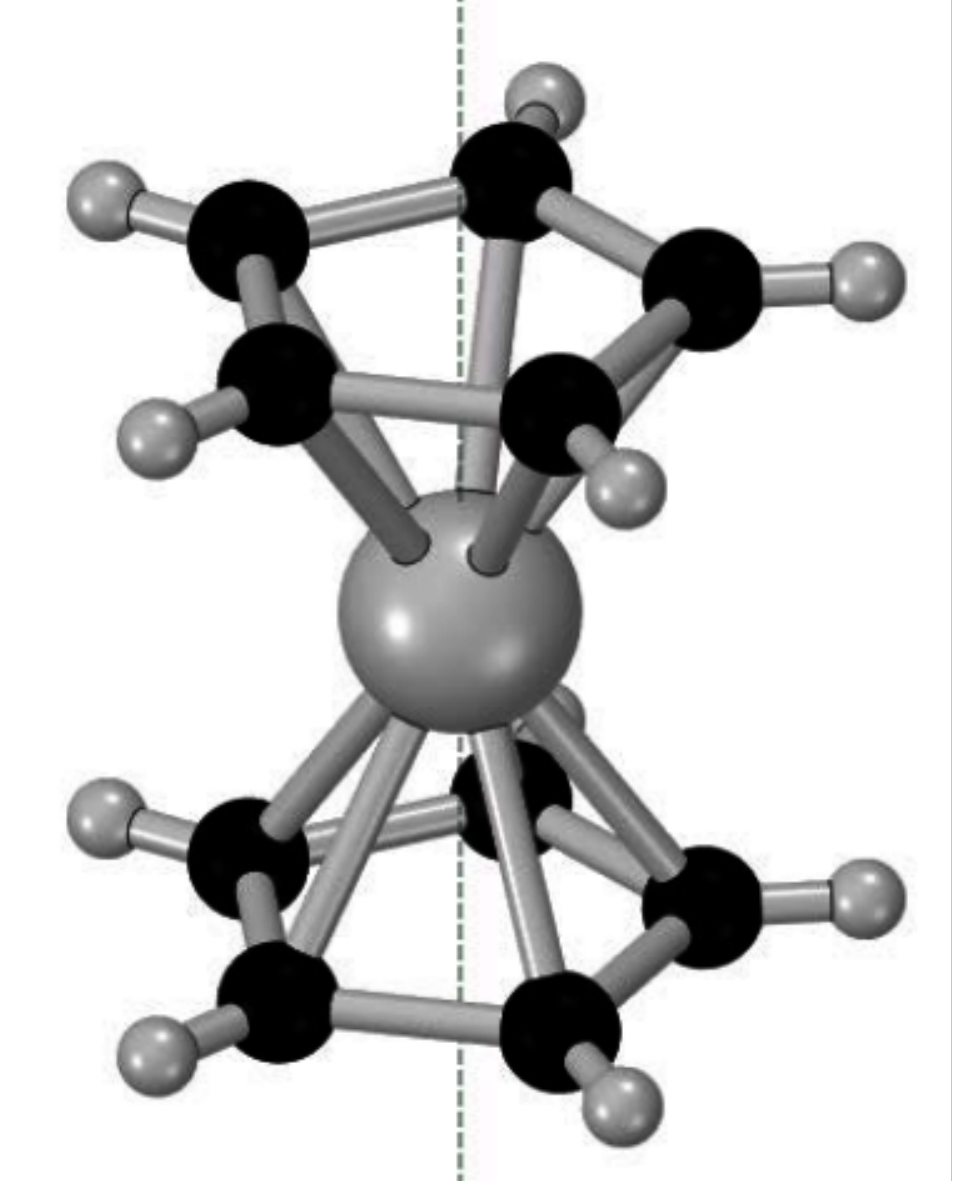}
  \caption{
    Model of nickelocene.
		Large gray, black, and small gray spheres represent Ni, C, and H atoms, respectively.
		A $C_5$ axis is indicated by a dashed line.}
		\label{scheme}
\end{figure}

We investigated nickelocene (denoted Nc below, Fig.~\ref{scheme}) adsorbed to an scanning tunneling microscope (STM) tip from a Cu(100) substrate, a system that has been characterized in a series of articles \cite{bac16, orm16, orm17a, orm17b}.
In particular, junctions with Nc  adsorbed to the tip were found to be stable at large conductances and transport calculations predicted that their electron transmission is spin polarized.

We used such tips to record spectra of the differential conductance from the tunneling range deep into contact.
Using density functional theory (DFT) calculations, we studied new molecular conformations and the corresponding degrees of spin polarization of the $d$ orbitals relevant from electron transport.
We also used a many-body Anderson impurity model to reproduce the evolution of the conductance spectra from spin excitation steps in the tunneling range to contact, where renormalization due to coupling to the conduction electrons leads to different line shapes.
Finally, our shot noise data reveal a noise reduction that is consistent with transport through spin-polarized molecular $d_{xz}$ and $d_{yz}$ orbitals.
In view of the DFT predictions on spin polarization, this observation excludes a horizontal orientation of the molecule at the tip.

\section{Prior results for Nc on Cu(100)}

Nc is comprised of two cyclopentadienyl rings that sandwich a Ni ion.
As discussed in Refs.~\onlinecite{bac16, orm16, orm17a, orm17b}, the molecule adsorbs upright on Cu(100), has a spin of 1, and exhibits a spin excitation that leads to steps in differential conductance ($dI/dV$) spectra at sample voltages $V\approx \pm 3$\,mV\@.
When a molecule is contacted with the STM tip, it is usually transferred to the tip.
These molecular tips are stable up to large conductances $G=I/V.$
At elevated $G$, $dI/dV$ spectra reversibly change to a peak at the Fermi level.
This spectral signature has been assigned to a spin-$\nicefrac{1}{2}$ Kondo effect, in agreement with the reduction of spin found by DFT simulating the experimental conditions~\cite{orm17a}.
Figure~4 of Ref.~\onlinecite{orm17a} shows the reduction of the molecular spin.
This reduction is due to the increase of electron density at the molecular junction and leads to an effective screening of the electron-electron interaction.
This reduces the spin polarization of the molecule, ultimately leading the spin 1 to $\nicefrac{1}{2}$ transition.
A similar change from spin excitations, which imply a magnetic anisotropy, to a Kondo effect was reported from substituted Fe porphyrins \cite{Karan_2017}.
In the present case, the transition appears abrupt and corresponds to a quick rise of the low-bias conductance.

According to DFT calculations, the molecule can attach to a single atom at the tip and get slightly tilted.
The simulated constant-current STM images using such a tip are in good agreement with the experimental ones~\cite{orm17a}.
Similar results have been reported for Nc-functionalized tips on Ag(110)~\cite{Wilson_2019} and Cu(111)~\cite{lliu}.
The tilt disappears as the tip is brought closer to the surface.
Transport calculations suggest that equal shares of the current are carried by the two molecular $d_{xz}$ and $d_{yz}$ orbitals, which are virtually degenerate at contact.
The calculated transmissions $T$ for Fermi-level spin-up and spin-down electrons, however, are different.
Upon contact formation, the spin polarization decreases from values above 70\,\% in the tunneling range to 33\,\% at $G=1.1$\,G$_0$ and 23\,\% at  1.6\,G$_0$.

\section{Experimental details}

We used an  ultra-high vacuum (UHV) low-temperature STM operated at 4.5~K\@.
The Cu(100) surface was prepared by Ar sputtering and annealing.
Tips were electrochemically etched from W wire and further prepared in UHV by annealing and indenting them into the Cu crystal to coat them with copper.
Finally, the sample was gently contacted until single Cu atoms were deposited from the tip and the contacts showed a conductance $G\approx 1\,\text{G}_0.$
Nc was evaporated onto the cold Cu sample from a Knudsen cell.
A Nc molecule was transferred from the substrate to the tip by bringing the tip closer to the center of the molecule until a rapid rise was observed in conductance-displacement data, $G(\Delta z),$ which were recorded simultaneously.

\section{Imaging of Nc tips}

\begin{figure}[hbt]
\includegraphics[width=0.89\columnwidth]{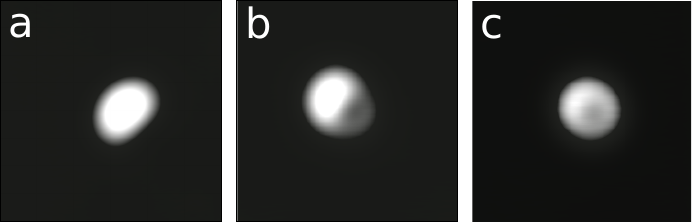}
\caption{(a--c) Constant-current topographs ($V=26$\,mV, $I=50$\,pA, (3\,nm)$^2$) of single Cu adatoms on Cu(111) recorded with different Nc tips.}
\label{topos}
\end{figure}

For characterization, we used the Nc tips to image Cu adatoms (Fig.~\ref{topos}) \cite{gsc}.
We typically observed the three patterns shown in panels a--c.
The tallest structure (maximum height 105\,pm, Fig.~\ref{topos}a) was found with 50\,\% of the Nc tips and appears approximately elliptical.
The pattern of Fig.~\ref{topos}b is slightly lower (90\,pm) and matches the image reported in Ref.~\onlinecite{orm17a}.
Approximately 40\,\% of our tips displayed this pattern.
Finally, Fig.~\ref{topos}c shows a nearly symmetric ''doughnut'' with a maximal height of 75\,pm.

The most simple model explaining these patterns is to assume that the Nc molecule can adsorb to the tip in different angles with the tip axis.
While the long axis of the molecule is strongly tilted in Fig.~\ref{topos}a, it is almost parallel to the tip in c.
This model matches the observed image symmetries and is also qualitatively consistent with the observed heights.

\begin{figure}[hbt]
  \includegraphics[width=0.99\columnwidth]{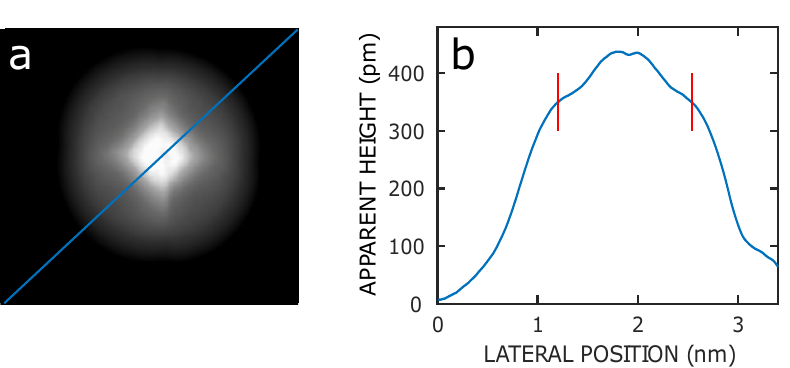}
  \caption{(a) Constant-current image of a Nc molecule at a Cu adatom recorded with a metal tip.
    The molecule had previously been transferred from a Nc-terminated tip to the adatom on the substrate.
    (b) Cross-sectional profile of the image in (a).
    The height of the shoulders (red markers) matches that of Nc  (350\,pm).
    The shoulder-shoulder distance (1.1\,nm) is twice the diameter of Nc in STM images.
    The maximal height (430\,pm) matches that of Nc plus that of a Cu adatom (80\,pm), which may be a coincidence.}
  \label{ncrot}
\end{figure}

Ref.~\onlinecite{orm17a} proposed that Nc at the tip adsorbs to a single apex atom via two C atoms of a cyclopentadienyl ring.
In a DFT calculation, this geometry was predicted for Nc at a single adatom on a Cu plane.
We tried to verify this structure by first transferring Nc from the surface to the tip and then depositing it on an adatom.
Figure~\ref{ncrot} shows that the resulting complex is not stable during STM imaging.
The constant-current image indicates that the Nc molecule is easily moved by the tip (or may be moving by itself) between four equivalent adsorption sites of the Cu(100) substrate that are centered around the adatom.
This observation suggests that the geometry used for the DFT calculations~\cite{orm17a}, namely a single atom on a Cu (100), may not be a realistic model to reveal the most stable adsorption geometry for Nc at the tip.

\section{$G(z)$ and $dI/dV$ data}

\begin{figure}[hbt]
  \includegraphics[width=0.99\columnwidth]{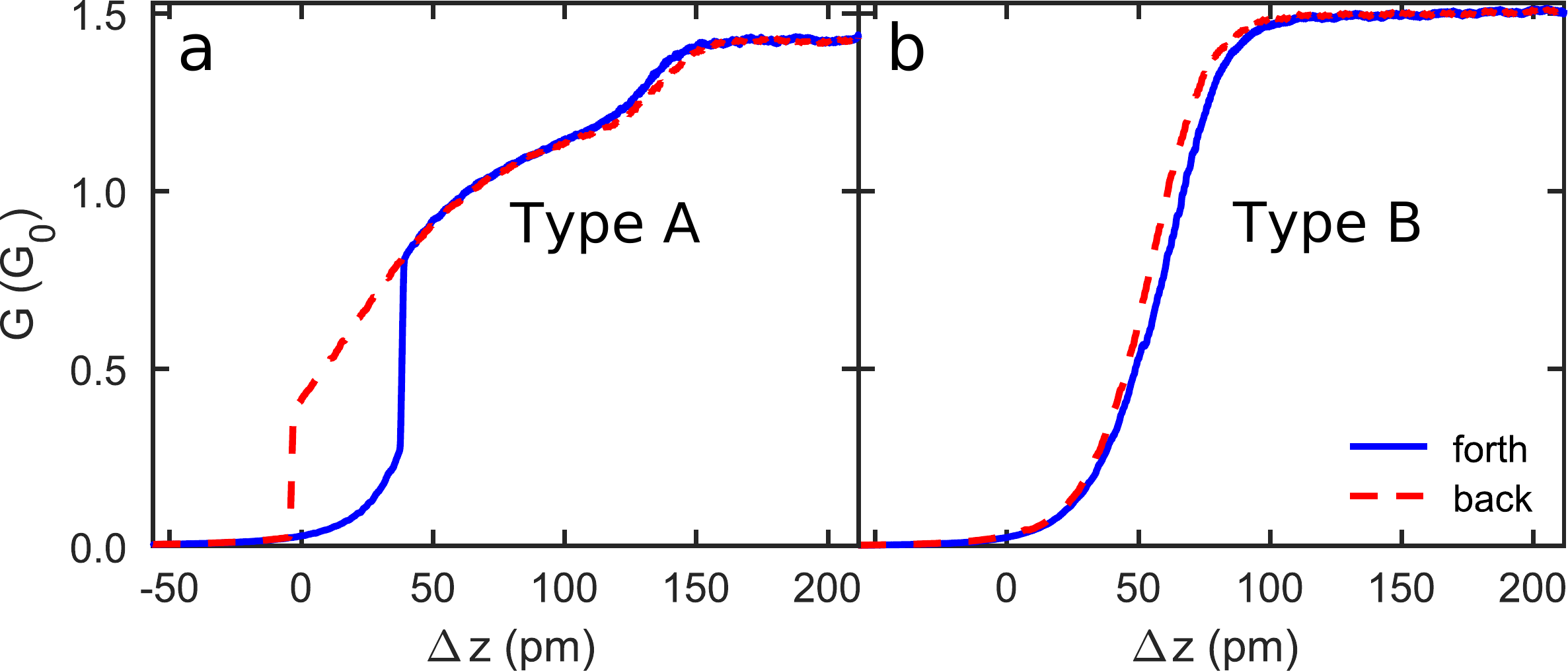}
  \caption{
    Conductance $G$ of nickelocene junctions vs.\ vertical displacement of the STM tip measured with $V=2$\,mV\@.
    $\Delta z=0$ corresponds to the tip position at $I=4.5$\,nA\@.
    Solid and dashed lines indicate approach and retraction of the tip.
    Two data sets \textbf{A} and \textbf{B} recorded with different tips on different molecules are shown in (a) and (b).
    \textbf{A} and \textbf{B} were observed in $\nicefrac{2}{3}$ and  $\nicefrac{1}{3}$ of the measurements, respectively.
    At small $\Delta z$, case \textbf{A} is similar to the data reported in Ref.~\onlinecite{orm17a}.
    Following an exponential rise at large separations ($\Delta z <0$, solid line), $G$ jumps to $\approx0.7$\,G$_0$.
    However, $G$ steadily increases at closer approach and saturates at $\approx 1.4$\,G$_0$ with an intervening rapid rise near $\Delta z \approx 130$\,pm.
    The conductance recorded while the tip was retracted (dashed line) exhibits hysteresis.
    In contrast, \textbf{B} exhibits no discernible hysteresis between the approach and retraction data.
    Both \textbf{A} and \textbf{B} reveal conductances of $\approx 1.4$\,G$_0$ when the tip is brought sufficiently close to the molecule.}
  \label{gz}
\end{figure}

The conductance of the Nc contacts varies in characteristic manners when the tip is brought closer to the sample.
>From previous conductance measurements of Nc, an abrupt rise to $G\approx0.7$\,G$_0$ was reported \cite{orm17a}.
We extended our measurements beyond this point toward smaller tip-molecule distances.
Two characteristic observations, labeled \textbf{A} and \textbf{B}, are shown in Fig.~\ref{gz}.
In case \textbf{A}, observed in approximately \nicefrac{2}{3} of all experiments, $G$ escalates to $\approx0.7$\,G$_0$ as the tip is brought closer from large separations ($\Delta z <0$, solid line), in agreement with earlier reports.
Further increase of $\Delta z$ leads to a steady increase of $G$, a rapid rise around $\Delta z \approx 130$\,pm, and a saturation of the conductance at $\approx 1.4$\,G$_0.$
Close inspection actually reveals a small reduction of $G$ at the largest $\Delta z$.
As the tip is retracted (dashed line), hysteresis is observed.
In addition to these strongly hysteretic $G(\Delta z)$ curves, we often (\nicefrac{1}{3} of our measurements) observed a smooth conductance evolution as shown in Fig.~\ref{gz} (case \textbf{B}).
We did not observe a clear correlation between the two cases \textbf{A} and \textbf{B} and the different imaging properties of the tip shown in Fig.~\ref{topos}.

Importantly, the conductance data at large $\Delta z$ are very similar in cases \textbf{A} and \textbf{B}.
In fact, almost all reproducible $G(\Delta z)$ data sets exhibited a contact conductance in the range from 1.2 to 1.6\,G$_0$.
We therefore measured the current noise in this conductance range.
Moreover, the weak variation of $G$ with $\Delta z$ in this range simplifies stable measurements.

\begin{figure}[hbt]
  \includegraphics[width=0.99\columnwidth]{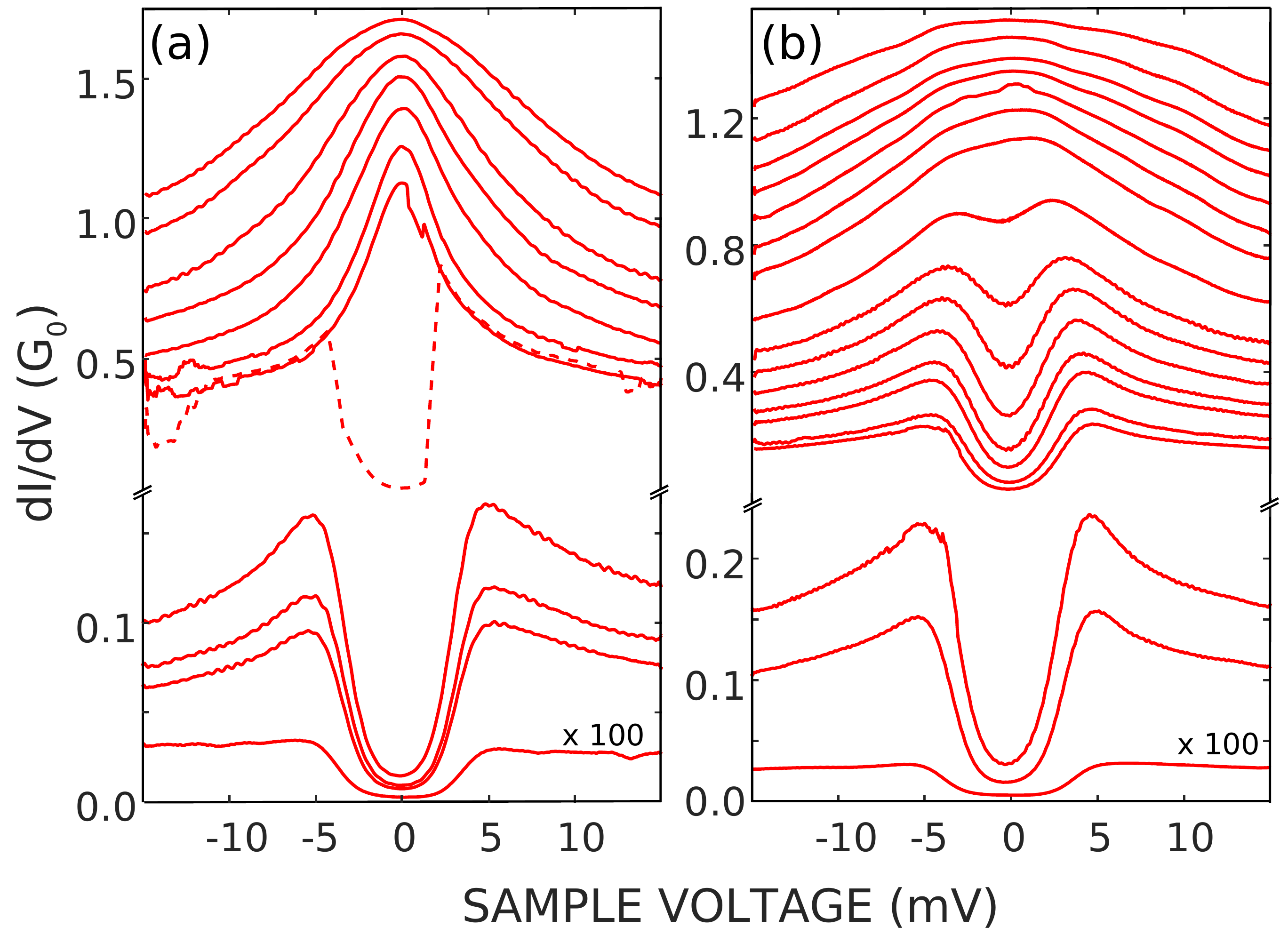}
  \caption{Spectra of the differential conductance $dI/dV$ of (a) \textbf{A} and (b) \textbf{B} Nc junctions measured from the tunneling into the contact range.
    The initial tip height was defined by a current $I=5$\,nA at $V=2$\,mV\@.
    The data in (a), lower part, were measured at tip displacements $\Delta z$ of $-100$, 5, 10, and 24\,pm.
    The upper data sets were recorded at $\Delta z = 42$ to 165\,pm.
    For clarity, the top four spectra were shifted upward in steps of 0.05\,G$_0$.
    During the measurement shown by a dashed line the junction abruptly changed from the peak shape at contact to the dip shape observed in the tunneling range.
    (b), lower part, shows data for $\Delta z$ of $-100$, 5, and 10\,pm.
    The upper part corresponds to $\Delta z= 24$ to 100\,pm.
    The top five spectra were shifted upward in steps of 0.05\,G$_0$.
    The bottom spectra in (a) and (b) were scaled up by a factor of 100.
		}
  \label{didv}
\end{figure}

Spectra of the differential conductance $dI/dV$ of our Nc junctions are displayed in Fig.~\ref{didv} for cases \textbf{A} and \textbf{B}\@.
In both cases, spectra in the tunneling range (bottom curves in Fig.~\ref{didv}) exhibit steps at $\pm 3.7$\,mV\@.
These steps have been assigned to spin transitions of Nc from $m_S=0$ to $1$ \cite{orm17b}.
The overshoots near $\pm5$\,mV may be understood both from non-equilibrium effects \cite{lot14,orm17a} and, as
we will see below (Sec.~\ref{sec:model}), from renormalization due to Kondo exchange coupling\cite{Jacob:PRB:2018}.
Clear differences between \textbf{A} and \textbf{B} exist at intermediate conductances (middle curves).
In case \textbf{A}, the spectra match those reported in Ref.~\onlinecite{orm17a}.
The spectral shape evolves into a single symmetric peak that has been assigned to a Kondo effect.
In the spectra of type \textbf{B} the overshoots gradually become more dominant until the conductance steps merge to a single feature.
Finally, at the largest conductances investigated (top curves), \textbf{A} and \textbf{B} display similar conductance data with a broad peak centered at zero bias.

The origin of the sharp transition found in \textbf{A} can be traced back to the change of spin in the molecule. A spin 1 presents a gap in the conductance, while the spin $\nicefrac{1}{2}$ presents a peak.
A fast switching from one to the other at a bias below the spin-flip threshold for spin 1 leads to a fast jump between a low conductance regime to a high one.
At larger bias, the change becomes very smooth (Supplemental information of Ref.~\onlinecite{orm17a}).

\section{DFT calculations}
\label{sec:dft}

Total-energy calculations to determine the molecular geometry at a single-atom-terminated tip were performed using the planewave code VASP~\cite{Kresse1996, Kresse1999} and for the transmission the localized-basis-set code Transiesta~\cite{Papior}.
The PBE exchange-and-correlation functional code~\cite{Perdew1996} was used with van-der-Waals D2 corrections~\cite{Grimme2006}.
The calculations closely follow the ones of Ref.~\onlinecite{orm17a}.
The novelty here is the exploration of new molecular conformations.

On top of a single atom, we find that the tilted conformation is only 5~meV more stable than an upright conformation.
This is in good agreement with the experimental observation that the molecule is unstable during imaging on a single atom on Cu(100) (Fig.~\ref{ncrot}).
As reported before~\cite{orm17a}, as the tip and surface approach, the spin of the molecule undergoes a sharp transition with interelectrode distance.
The spin is $S=1$ for a molecule adsorbed at the tip, and becomes $S=\nicefrac{1}{2}$ in the molecular junction after the transition.
Interestingly, when the molecule is adsorbed parallel to the surface, the molecular spin is computed to be 1, down to a small distance between electrodes.

\begin{figure}[hbt]
\includegraphics[width=0.99\columnwidth]{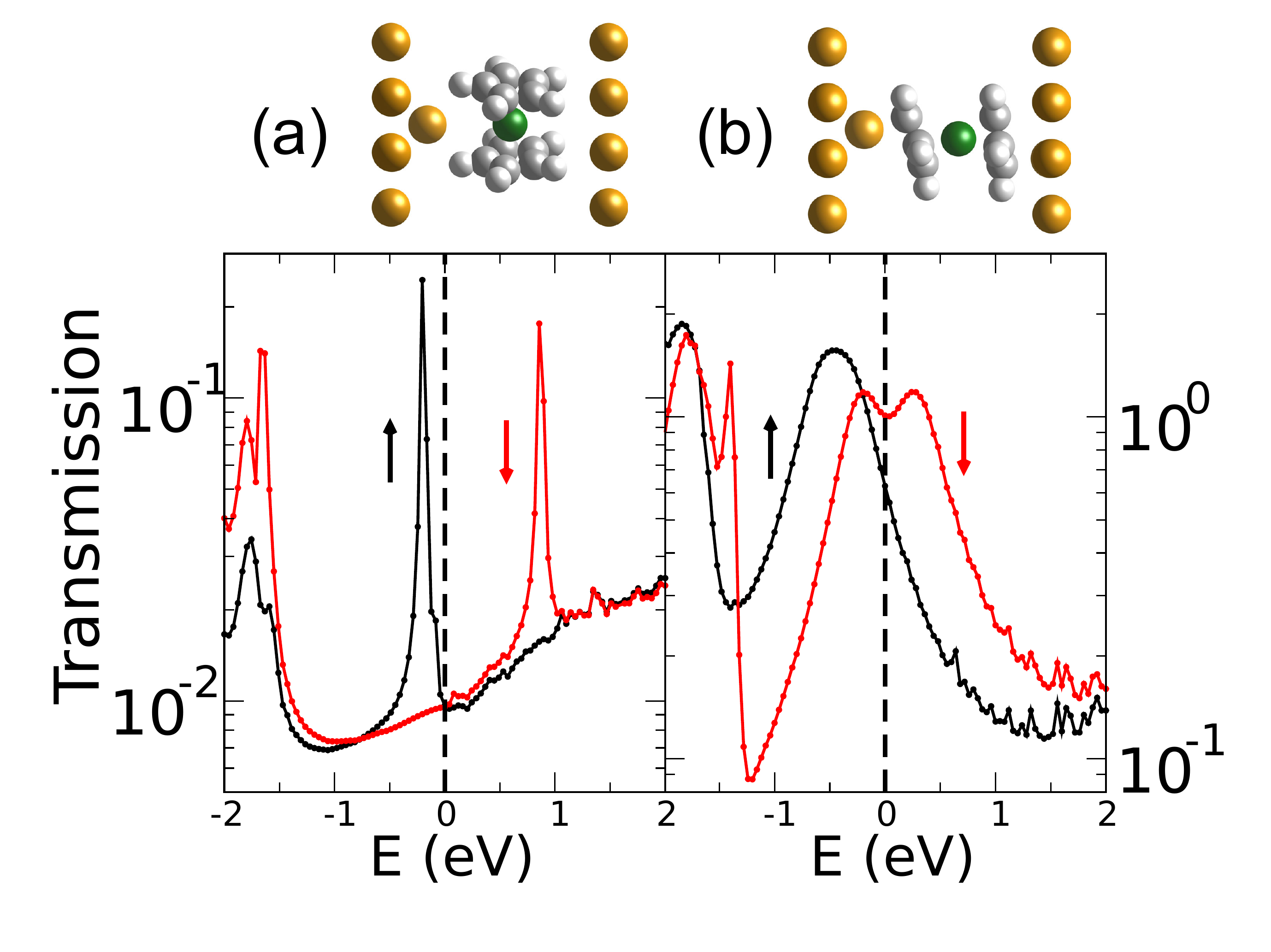}
\caption{
Calculated structures and corresponding spin-polarized electron transmissions (up and down arrows correspond to spin up and down, respectively).
In (a) the molecular axis is parallel to the planes of the electrodes with an interplane distance of 9.96~\AA, and in (b) perpendicular to the planes with an interplane distance of 10.16~\AA\@.
The vertical configuration, (b), yields a strong and distinct spin polarization while the parallel one, (a),  gives a negligible one.}
\label{transmission}
\end{figure}

When the molecule is adsorbed on a tip atom with its axis parallel to the surface, the energy is 40\,meV lower than for the tilted conformation.
As the tip approaches the surface, the surface starts affecting the molecule and destabilizes the parallel
conformation with respect to the vertical one.
Our calculations show that the parallel conformation depicted in Figure~\ref{transmission}(a)
is 900\,meV higher in energy than the vertical arrangement of Figure~\ref{transmission}(b).
This probably implies that vertical configurations are more likely when the tip and surface are brought together.

Figure~\ref{transmission}(a) shows the electron transmission for a molecule with its axis parallel to the surface, and a tip formed by a Cu (100) surface and one adatom, at a distance of 9.96~\AA\@.
In (b) the transmission is shown for the vertical configuration.
In the latter case the electron transmission at the Fermi level is dominated by the minority molecular spin.  This leads to a sizeable spin polarization of the electron transport.
Two channels are shown to contribute to the transmission as expected from the degenerate molecular orbitals
of $d_{xz}$ or $d_{yz}$ characters.
Interestingly, the parallel configuration yields negligible spin polarization at the Fermi level.
This is due to the smaller interaction of the molecule with the electrodes that reduce the contribution of $d_{xz}$ or $d_{yz}$ to the transmission at the Fermi level.
Shot noise measurements are very sensitive to the spin-polarization of the transmission and should be able to discriminate between these two limiting cases.

The spin-polarized transmission is due to the open-shell electronic structure and finite spin of the molecule.  This polarization does not require a fixed orientation of the molecular spin, whose direction likely is fluctuating at the temperature of the experiment.
In spite of the polarization of the transmission the current consequently does not need to be spin-polarized.

\section{Modeling of $dI/dV$ spectra}
\label{sec:model}

The inelastic conductance can be modelled by using many-body theory.
For a general account of inelastic tunneling spectra, we refer the reader to Refs.~\onlinecite{PSS2012, Markus2017}.
For modeling experimental spectra of the differential conductance we focus on Fig.~\ref{didv}(b).
The tips leading to these spectra did not show abrupt changes in the conductance-distance data (Fig.~\ref{gz}). 
According to the DFT calculations the spin-1 of the Nc molecule responsible for the spin excitations is localized in the $d_{xz}$- and $d_{yz}$-orbitals of the Ni center, which are nearly degenerate.
We hence model the system by a two-orbital Anderson impurity and include a magnetic anisotropy term \cite{Jacob:PRB:2018,Li:NL:2019}.
The impurity part of the Hamiltonian is then given by:
\begin{eqnarray}
  \label{eq:model}
  {\cal H}_{\rm imp} &=& \epsilon_d \hat{N}_d
  + U \sum_{\alpha} \hat{n}_{\alpha\up} \, \hat{n}_{\alpha\dn}
  + U^\prime \sum_{{\alpha,\alpha^\prime}\atop{\alpha\neq\alpha^\prime}} \hat{n}_{\alpha} \, \hat{n}_{\alpha^\prime}
  \nonumber\\
  && - J_{\rm H} \sum_{{\alpha,\alpha^\prime}\atop{\alpha\neq\alpha^\prime}} \vec{S}_\alpha \cdot \vec{S}_{\alpha^\prime}
  + D \hat{S}_z^2.
\end{eqnarray}
where $\hat{N}_d=\sum_{\alpha,\sigma}\hat{n}_{\alpha\sigma}$ is the number operator for both $d$-levels $\alpha=1,2$, $\hat{n}_{\alpha\sigma}=d_{\alpha\sigma}^\dagger d_{\alpha\sigma}$ is the number operator of an individual $d$-level $\alpha$ with spin $\sigma$, $U$ is the intra-orbital, $U^\prime$ the inter-orbital Coulomb repulsion, $J_{\rm H}$ the Hund's coupling, $\vec{S}_\alpha$ measures the total spin of an individual $d$-level $\alpha$, \ie\ $\vec{S}_\alpha= \sum_{\sigma\sigma^\prime} d_{\alpha\sigma}^\dagger \vec\tau_{\sigma\sigma^\prime} d_{\alpha\sigma^\prime}$ with $\vec{\tau}={}^t(\tau_x,\tau_y,\tau_z)$ being the vector of Pauli matrices $\tau_i$, and $S_z$ is the $z$ component of the total spin of both impurity levels.

The crystal field in conjunction with the spin-orbit coupling gives rise to magnetic anisotropy \cite{Abragam:book:2012} which is taken into account by the effective spin Hamiltonian given by the last term of (\ref{eq:model}), where $D$ is the uniaxial anisotropy \cite{Gatteschi:book:2006}.
Here we use $D>0$ as experimentally found~\cite{orm16}.
Hence the degeneracy of the $S=1$ triplet is partially lifted, the $\ket{m_z=0}$ singlet state becomes the ground state, and the $\ket{m_z=\pm1}$ states an excited doublet.
Due to the lack of ground state degeneracy the Kondo effect is suppressed \cite{Jacob:PRB:2018}.

Recent calculations have shown that Kondo temperatures obtained under non-equilibrium and equilibrium conditions coincide.
We thus expect that in our description of Kondo physics here non-equilibrium features may be neglected \cite{nico2016}.
The two-orbital Anderson impurity model is solved within the one-crossing approximation which consists in a resummation of a subset of diagrams in the perturbation expansion in the coupling $\Gamma=\Gamma_\tip+\Gamma_\sub$ to the conduction electrons in the tip ($\Gamma_\tip$) and substrate ($\Gamma_\sub$) to infinite order \cite{Haule:PRB:2001}.
Details of the method can be found in previous works \cite{Jacob:EPJB:2016,Jacob:PRB:2018}.
The solution yields the spectral function $A_d(\omega)$ of the impurity levels.

\begin{figure}
  \includegraphics[width=\linewidth]{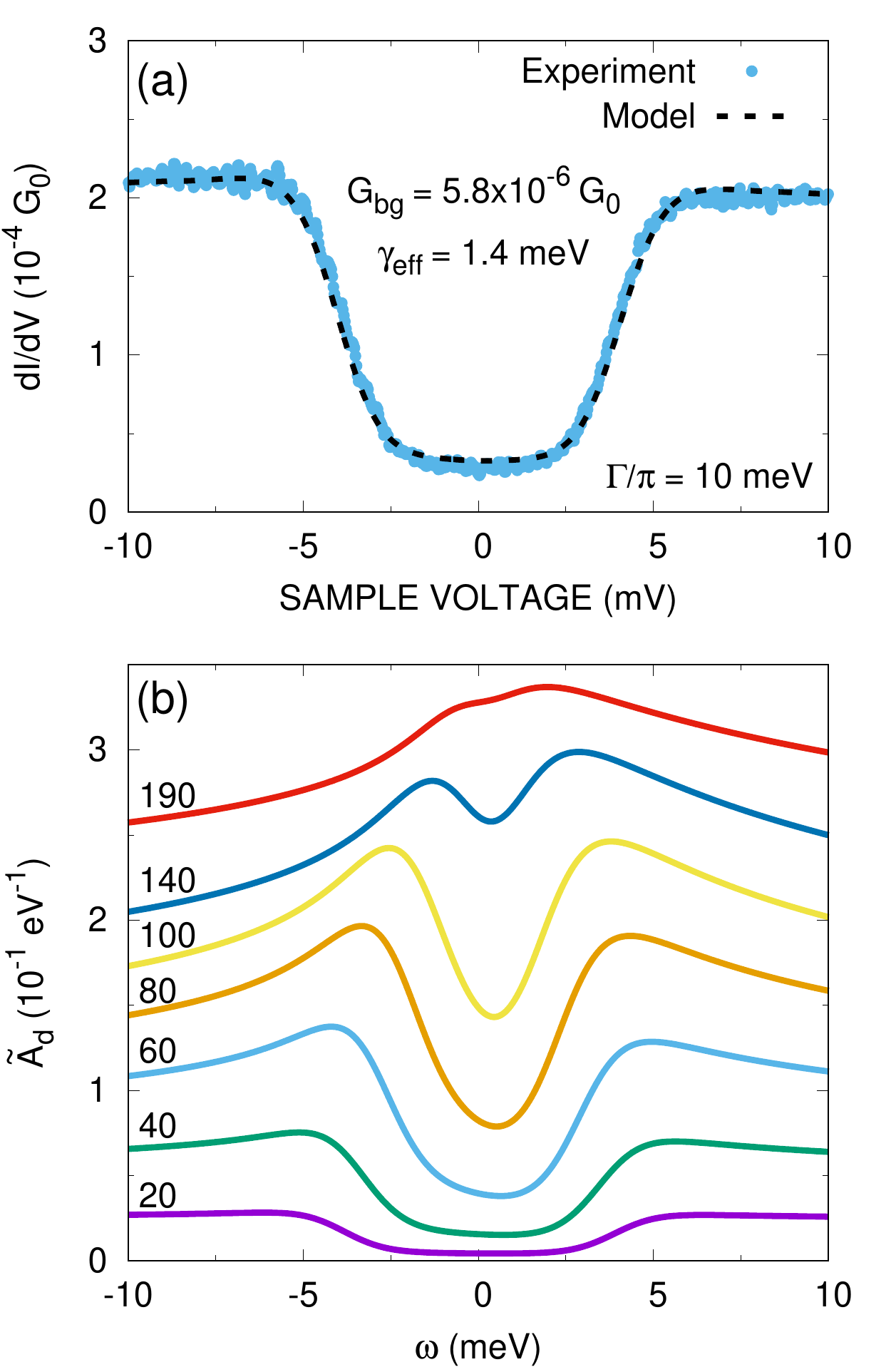}
  \caption{\label{fig:model_spectra}
   Results of two-orbital Anderson impurity model (\ref{eq:model}) calculations. 
   (a) Theoretical (dashed line) and experimental $dI/dV$ in the tunneling regime.
   The theoretical $dI/dV$ was obtained from the spectral function $\tilde{A}_d(\omega)$ for
   $\Gamma/\pi=10$~meV at $T\sim4.5$~K and fitting to the experimental $dI/dV$ according to Eq.~(\ref{eq:dIdV}). 
    (b) Temperature-smeared spectral functions $\tilde{A}_d(\omega)$ for different couplings $\Gamma/\pi$ (in meV) to the conduction electrons, calculated at $T\sim4.6$~K\@.
    Anderson model parameters: $U=3.5$~eV, $U^\prime=2.5$~eV, $J_H=0.5$~eV, $\epsilon_d=-3$~eV, and $D=4.2$~meV\@.}
\end{figure}

From the spectral function the $dI/dV$ spectra can be calculated as follows.
When the molecule is at the STM tip the applied bias $V$ essentially drops between the molecule and the substrate.
According to Ref.~ \onlinecite{Meir:PRL:1992}, we can then write the current through the two Ni $3d$-orbitals as:
\begin{equation}
  \label{eq:IV}
  I(V) = 4\pi \gamma_\eff \frac{2e}{h} \int d\omega \left[ f(\omega-eV)-f(\omega) \right] A_d(\omega),
\end{equation}
where we have assumed that the coupling to the leads is the same for both orbitals, and the factor
of 4 comes from summing over spin and orbital degrees of freedom.
$\gamma_\eff$ is the effective coupling of the impurity shell to the tip and the substrate given by
$\gamma_\eff = 2\,\Gamma_\sub\Gamma_\tip/(\Gamma_\sub+\Gamma_\tip)$ \footnote{
Note that $\gamma_\eff$ if defined in terms of the full widths caused by electrode couplings in Ref.~\onlinecite{Meir:PRL:1992}.
In the present paper, $\Gamma_\sub$ and $\Gamma_\tip$ are the corresponding half-widths, which leads to the factor of two in our definition.}.
The Fermi functions in Eq.~\ref{eq:IV} are those of the substrate (at bias $V$) and of the STM tip, respectively.

From Eq.~\ref{eq:IV} we obtain the differential conductance:
\begin{equation}
  \label{eq:dIdV}
  	\frac{dI}{dV} = 4\pi\, \gamma_\eff {\mathrm G_0}  \, \tilde{A}_d(eV) + G_{\bg},
				\end{equation}
where $\tilde{A}_d(eV)$ is the temperature smeared spectral function given by the convolution
\begin{equation}
  \label{eq:convolution}
  \tilde{A}_d(\omega) = \int d\omega^\prime\, (-f^\prime(\omega^\prime))\,A_d(\omega+\omega^\prime).
\end{equation}
We added a constant background conductance $G_{\bg}$ in Eq.~(\ref{eq:dIdV}) to describe the conductance via other molecular channels as well as direct tunneling from the tip to the surface.

We next choose the Anderson model parameters $D,\epsilon_d,\Gamma$ such as to match the experimental spectra in the tunneling regime, while the interaction parameters remain fixed ($U=3.5$~eV, $U^\prime=2.5$~eV and $J_H=0.5$~eV).
We find $D=4.2$~meV, $\epsilon_d=-5$~eV and $\Gamma/\pi=10$~meV\@.
The effective coupling $\gamma_\eff$ and the background conductance $G_{\bg}$ can now be determined for the tunneling regime by linear regression using Eq.~\ref{eq:dIdV}.
Fig.~\ref{fig:model_spectra}(a) shows the resulting differential conductance in the tunneling regime.
The agreement between the fitted model spectrum and the experimental one in the tunneling regime is quite remarkable. 
The slight asymmetry between the left and the right step is well reproduced and can therefore be attributed to charge fluctuations induced by the deviation from particle-hole symmetry as discussed in previous work \cite{Jacob:PRB:2018}.
The asymmetry is enhanced and eventually reversed as the tip is brought closer to the sample [Fig.~\ref{fig:model_spectra}(b)].

Fig.~\ref{fig:model_spectra}(b) displays the temperature broadened ($\approx 4.5$~K) spectral functions $\tilde{A}_d(\omega)$ computed for the model parameters found in the tunneling regime, but for different couplings $\Gamma=\Gamma_\tip+\Gamma_\sub$.
Again the asymmetry of the spectra stems from charge fluctuations induced by the deviation from particle-hole symmetry.
The calculated spectra qualitatively reproduce the evolution of the experimental data of Fig.~\ref{didv} as the molecule-tip system approaches the surface and the conductance increases.
The spin excitation energies at both polarities move inwards to lower energies.
Simultaneously the excitation steps, which initially are flat above the threshold, acquire an increasing overshoot.
At very large $\Gamma$ the gap finally begins to close as the step features become increasingly peak-like and merge.
Hence both the decrease in the spin excitation energies and the overshoot of the step features with increasing conductance observed in the experiments can be understood in terms of renormalization by Kondo exchange coupling with the conduction electrons, similar to previous works \cite{Oberg:NNano:2014, Jacob:PRB:2018, Li:NL:2019}.

In our calculations within the one-crossing approximation (OCA), further lowering of the temperature splits the single peak features again (not shown).
This was also observed in previous work \cite{Jacob:PRB:2018} and appears to be in agreement with numerical renormalization group (NRG) calculations for the single-channel spin-1 Kondo  model with positive uniaxial magnetic anisotropy that show a split-peak \cite{Zitko:PRB:2008, Schiller:PRB:2008}.
However, recent NRG calculations have shown that the splitting should actually not occur for the two-channel case considered here (two impurity levels coupled to two conduction electron baths) and thus is an artifact of the one-crossing approximation. 
Hence the formation of the peak feature does signal the onset of the Kondo effect and marks a quantum phase transition \cite{Blesio:PRB:2019}.

In summary, the model calculations show that $dI/dV$ spectra recorded for systems which show the type \textbf{B} behavior in the conductance (Fig.~\ref{didv}b) are in agreement with a spin-1 molecule with the spin localized in two degenerate $d$-orbitals subject to positive uniaxial magnetic anisotropy coupled to electrodes.
The evolution of the spectra as the tip-molecule system approaches the substrate can largely be understood by renormalization due to exchange coupling with the conduction electrons \cite{Oberg:NNano:2014, Jacob:PRB:2018}.
Another important contribution probably comes from non-equilibrium population of excited states \cite{lot14,orm17a,orm17b}, which is not captured here, since it would require an out-of-equilibrium many-body treatment of the system.

\section{Noise data}

Current noise was measured with a setup described in Ref.~\onlinecite{abu}.
Briefly, after preparing a contact the STM control electronics was disconnected and a battery-driven circuit was used to drive current through the junction.
The voltage noise present at series resistors was amplified by two parallel amplifiers and cross-correlation of the voltage signals was used to reduce amplifier noise.
The current noise was obtained from the voltage noise using the measured conductance.
The resulting spectrum of the noise power density was averaged over the range from 20~kHz and 200~kHz excluding spurious signals.
The upper limit reflects a low pass characteristic due to cable capacitances while the lower boundary was used to minimize the influence of $1/f$ noise.

\begin{figure}%%[hbt]
  \includegraphics[width=0.999\columnwidth]{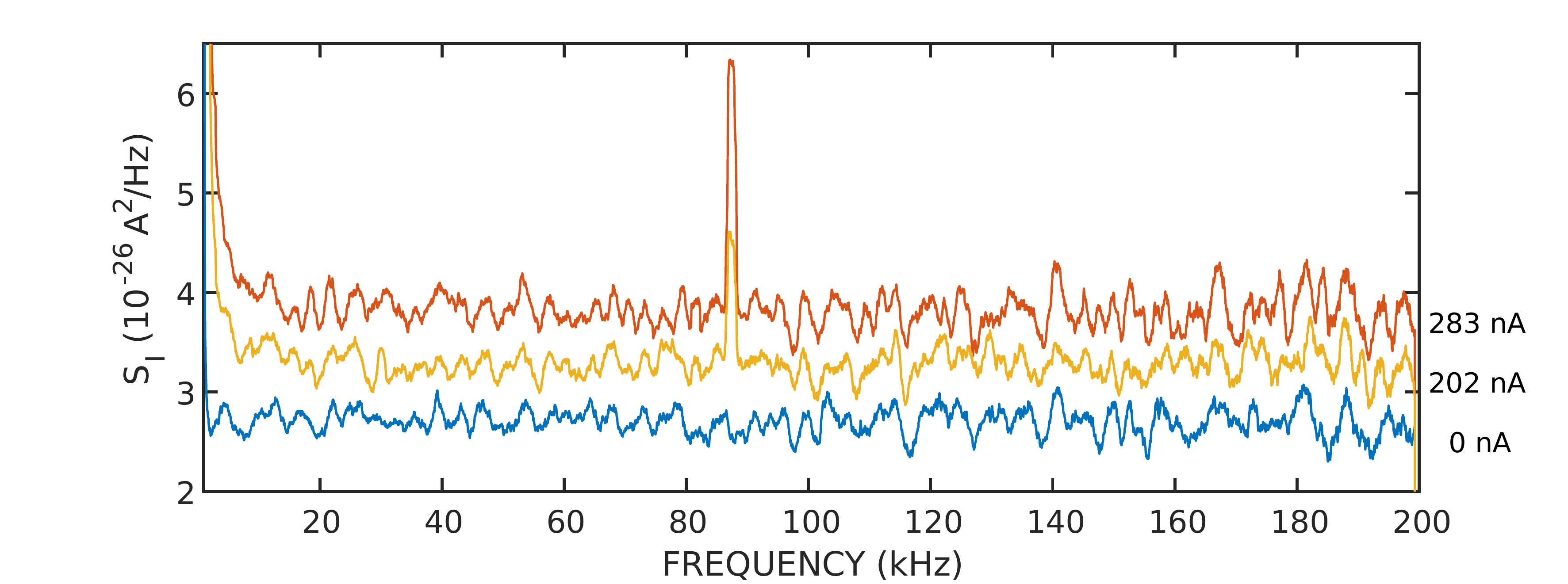}
  \caption{ Current noise density corrected for frequency response for bias currents of 0 (lowest spectrum), 202, and 283\,nA\@.
    A moving average filter has been applied to simplify comparison.
    The peak near 87\,kHz is due to electrical interference.
    $1/f$ noise is discernible at low frequencies in the 202 and 283\,nA data.}
  \label{sf}
\end{figure}

\begin{figure}%[hbt]
  \includegraphics[width=0.89\columnwidth]{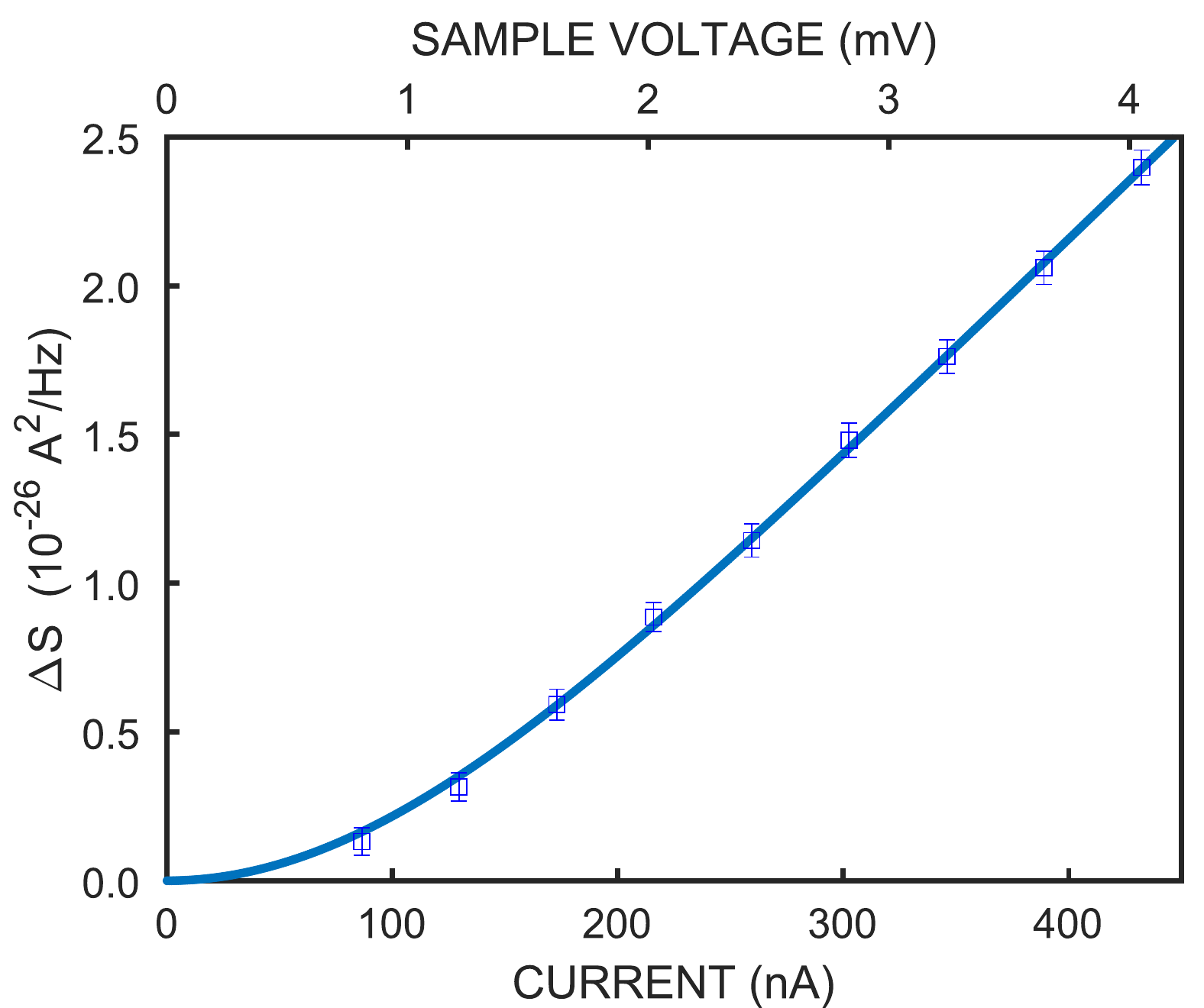}
  \caption{
    Excess noise power $\Delta S(I)$ measured on an isolated nickelocene molecule.
    Dots show experimental data, a line displays a fit using Eq.~\ref{leso}.
    The upper abscissa displays junction voltages.
    While the uncertainty of the noise power averaged of a 20\,kHz interval is indicated, the uncertainty of the current values is smaller than the symbol size.}
  \label{exc}
\end{figure}

To characterize a single contact, measurements were performed for several current values.
In addition to the noise, we also determined the DC voltage drop over the contact to calculate its conductance.
We only used data from contacts whose conductance remained constant throughout the entire set of measurements (typically 9 current values).
For comparison, we determined the conductance from the thermal noise $S_\theta = 4 k_B \theta G$ ($\theta$ temperature, $k_B$ Boltzmann constant) measured at $I=0$.
Both conductances agreed within $\pm 5$\,\%.

Spectra of the current noise $S$ \footnote{
 We define the power spectral density $S$ as the zero frequency limit of $2 \int_{-\infty}^{\infty} \mathrm{dt}\, e^{i \omega t} \left< \Delta I(0) \Delta I(t) \right> $, where $\left< ~ \right>$ indicates an ensemble average.
 A formal definition may be found in Ref.~\onlinecite{Kogan}}  
were measured with molecular tips on flat Cu(100) areas.
Figure \ref{sf} shows the increase of the spectral noise power density with the bias current through the Nc junction.
The excess noise $\Delta S$ was obtained using \cite{Lesovik, abu}
\begin{equation}
\Delta S = S - S_\theta = F\left[ S_0 \, \coth\left( \frac{S_0}{S_{\theta}}\right)-S_{\theta} \right],\label{leso}
\end{equation}
where $F=\lim_{\theta\ra0}\Delta{S}/S_0$ is the Fano factor and $S_0 = 2eI$ is the classical shot noise power density.
Equation \ref{leso} was used to fit the measured noise $S(I)$ treating $F$ and $\theta$ as adjustable parameters.
Figure \ref{exc} shows an example experimental data (dots) and the fit (line) obtained.
As expected for all but the lowest currents, the relation between $\Delta S$ and the current $I$ is almost linear.

\begin{figure}%[hbt]
  \includegraphics[width=0.99\columnwidth]{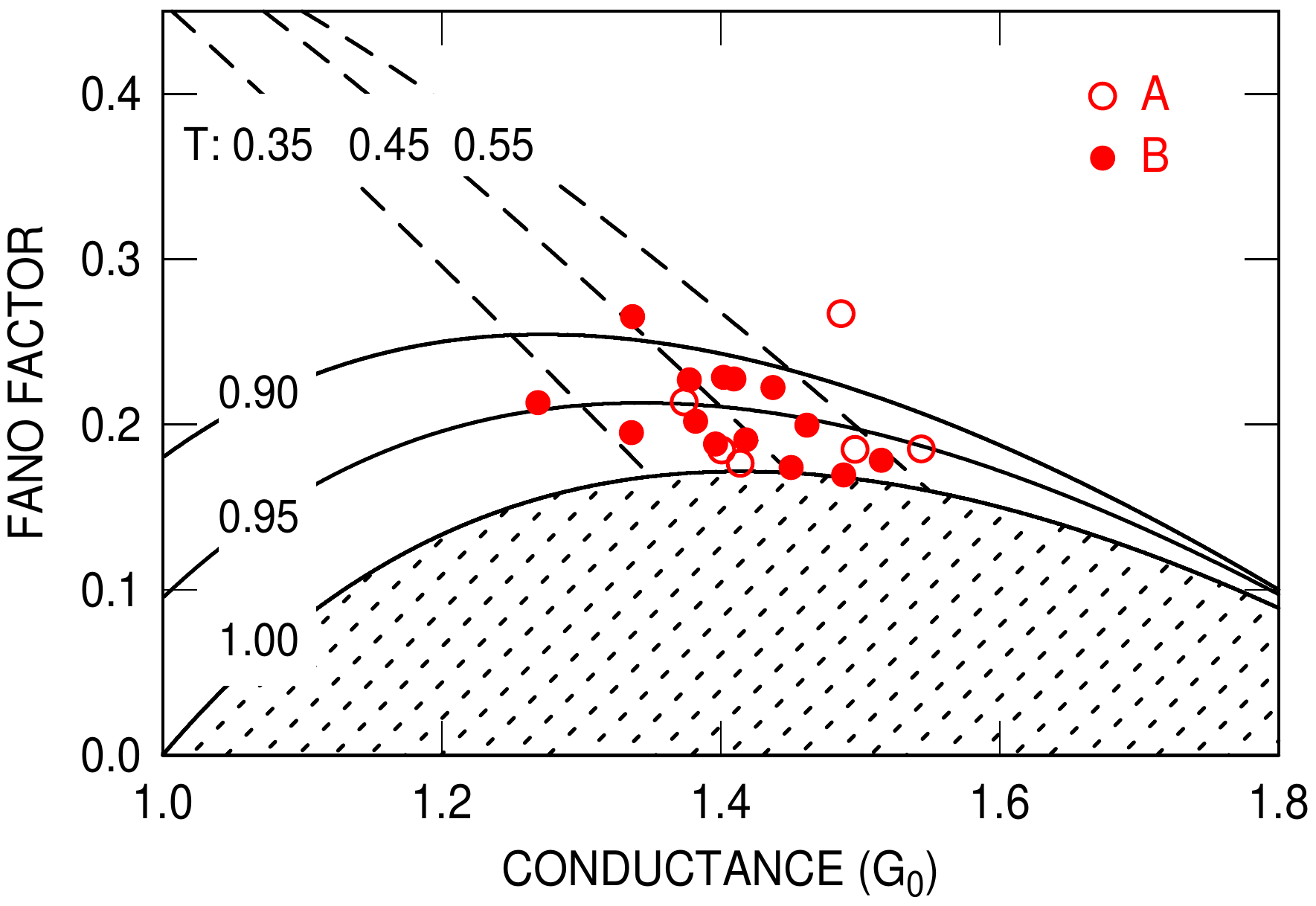}
  \caption{Fano factor vs.\ conductance of nickelocene junctions.
    Circles and dots indicate data from junctions with $G(z)$ curves of types \textbf{A} and \textbf{B}, respectively.
    No significant difference of the Fano factors in these cases is discernible.
    Solid and dashed lines show calculated Fano factors in the model of Eq.~\ref{twoch}, which assumes two identical transport channels, each comprising two spin sub-channels.
		The lines are labeled with the transmission $T$ of one of these sub-channels. 
		Combinations of $F$ and $G$ in the hatched area are inconsistent with the model. }
  \label{fano}
\end{figure}

Figure \ref{fano} shows the obtained Fano factors vs.\ the junction conductances.
While there is some scatter, the data are confined to the conductance range 1.3 to 1.5~G$_0$ and Fano factors vary between 0.17 and 0.27.
The arithmetic means are $\langle G \rangle =1.42$\,G$_0$ and $\langle F \rangle =0.20.$

To analyze the Fano factor we use a Landauer-B\"uttiker description \cite{blanter},
\begin{equation}
F= \frac{ {\mathrm G_0}}{2 G} \sum_i{T_i(1-T_i)},
\label{fallg}
\end{equation}
where the sum extends over all spin-resolved transport channels with transmissions $T_i$. 

Fig.~\ref{transmission}(b) shows that the current at contact is essentially carried by two nearly degenerate molecular orbitals, $d_{xz}$ and $d_{yz}$.
The transmissions of these orbitals have been predicted to be identical but strongly spin dependent \cite{orm17a}.
We therefore assume that the current is carried by two identical channels, each comprising sub-channels with transmissions $T_\uparrow$ and $T_\downarrow$ for up and down spins, respectively.
Equation \ref{fallg} then reduces to 
\begin{equation}
F= \frac{\mathrm G_0}{G} \, [ T_\uparrow (1-T_\uparrow) + T_\downarrow (1-T_\downarrow) ].
\label{twoch}
\end{equation}

Figure~\ref{fano} shows Eq.~\ref{twoch} evaluated for a range of conductances $G$ and a few transmission values $T\uparrow$.
$T_\downarrow$ was adjusted to obtain the average experimental conductance, $T_\downarrow + T_\uparrow = G/{\mathrm G_0}=1.4$.
Within this model, a fit of the experimental data is obtained for the transmissions $T_\downarrow \approx 0.45$ and $ T_\uparrow \approx 0.95$ corresponding to a spin polarization $(T_\uparrow-T_\downarrow)/(T_\uparrow+T_\downarrow) \approx 36 \%.$
This experimental result is in reasonable agreement with the spin polarization calculated for a nearly vertical molecule (33\,\% and 23\,\% at $G=1.1$ and 1.6\,G$_0$, respectively) \cite{orm17a}.
A horizontal orientation, however, is predicted to exhibit no spin polarization and can therefore be ruled out.
It should be noted that both $d$ channels exhibit identical transmissions according to our DFT calculations.
If we allow for different transmissions $T(d_{xz}) \neq T(d_{yz})$ the spin polarization extracted from the Fano factor is reduced.

The above analysis of the measured shot noise in terms of the Landauer-B\"uttiker picture yields a mean-field description of the correlated transport via the molecular quantum spin.
It can therefore not be expected to give a correct description in the Kondo regime, for which deviations from the Landauer-B\"uttiker description have been predicted \cite{sel6, vit8, mor8, coc19}.
Unfortunately, the deviations between the experimental data and the fit observed in Fig.~\ref{exc} are too small to test these predictions.

\section{Conclusions}

Systematic studies of the electronic conductance as a Nc-terminated tip approaches a Cu(100) substrate reveal that there are two types of molecular tip.
Type \textbf{A} leads to a first abrupt change in conductance close to G$_0$, later on a second fast rise, and finally conductances near 1.4\,${\mathrm G_0}$, for very short junctions.
Type \textbf{B} rather shows a smooth evolution from tunneling to contact with a plateau at conductances similar to 1.4\,${\mathrm G_0}$.
These two classes of tips show different $dI/dV$ spectra at different tip-surface distances.
Type \textbf{A} exhibits an abrupt transition from symmetrical steps at positive and negative bias reflecting spin excitation thresholds to a zero-bias anomaly typical of a Kondo peak~\cite{orm17a}.
Type \textbf{B} displays instead a gradual closing of the inelastic gap.

Our calculations show that different molecular conformations can explain different spin states of the junctions. 
Type \textbf{A} can be associated with a nearly vertical configuration of the molecule where the electrodes contact the
cyclopentadienyl moieties.
The behavior of type \textbf{A} is consistent with an abrupt spin 1 to \nicefrac{1}{2} transition caused by electronic screening of the molecular charging energy~\cite{orm17a}.
Type \textbf{B} is more difficult to assign to a given geometry.
%We find that molecules lying parallel to the electrodes keep their spin of 1 up to very small electrode separations.

As to the deep contact regime, the similar conductance spectra and shot noise data from \textbf{A} and \textbf{B} tips hint that the molecule is in similar environments with both kinds of tip.
The molecule is more rigid than the tip which consequently may suffer a deformation.
In a simple-minded picture, the molecule finds itself close to many metal atoms, somewhat independent of the initial conditions (\textbf{A} or \textbf{B} tip).
This picture leads to a final screened state of the molecule that likely exhibits $S=\nicefrac{1}{2}$.
The shot-noise measurements indeed show that the transmission is spin polarized in the deep contact range.
This is in good agreement with our mean-field Landauer-B\"uttiker calculations for the $S=\nicefrac{1}{2}$ state.
The corresponding spin polarization is close to the one predicted by the calculations at approximately 35\%.

On the other hand, our many-body calculations enable modeling of the conductance spectra and show that a smooth transition from inelastic steps to a broad zero-bias peak as observed for \textbf{B} tips is also possible via the renormalization of the magnetic excitations of a $S=1$ molecule.
The available experimental data are therefore compatible with both theoretical scenarios that predict a conductance peak at zero bias and $T=4.5$\,K\@.

%Landauer-B\"uttiker calculations for the $S=1$ state have not been performed.
%However, a spin-polarized transmission is also expected in a $S=1$ scenario that therefore cannot be excluded.

\section{Acknowledgements}

We thank Laurent Limot for sharing experimental details with us, Dirk K. Morr for sending a manuscript prior to publication, and Jonas Fransson and Armando A. Aligia for discussions.
This work has been supported by the Deutsche Forschungsgemeinschaft, project BE 2132/8.
NL is grateful to financial support from MICINN (RTI2018-097895-B-C44).
DJ acknowledges funding by the grant “Grupos Consolidados UPV/EHU del Gobierno Vasco” (IT1249-19).

%\bibliography{NiOc,Noise,theory}

\end{document}